\documentclass[preprint,showpacs,floats,letterpaper,floatfix,,groupedaddress,eqsecnum]{revtex4}

\usepackage{subfigure}
\usepackage{amssymb,amsmath}
\usepackage[dvips]{graphicx}
\usepackage{csquotes}
\usepackage{enumerate}
\usepackage{dcolumn,epsfig}
\usepackage[left]{lineno}
\usepackage{ifthen,color,fancyvrb}
\usepackage{ulem}

\begin{document}

\title{Theoretical and numerical study of vibrational resonance in a damped softening Duffing oscillator}

\author{Ivan Skhem Sawkmie$^1$, and Donrich Kharkongor$^2$}
\email{dkharkongor@anthonys.ac.in}

\affiliation{$^1$Department of Physics, North-Eastern Hill University,
Shillong-793022, India}
\affiliation{$^2$Department of Physics, St. Anthony's College,
Shillong-793003, India}

\begin{abstract}
We study the possibility of occurrence of vibrational resonance in a
softening Duffing oscillator in the underdamped and overdamped cases 
both theoretically as well as numerically. The oscillator is driven 
by two periodic forces. Numerically we find that in the underdamped case two oscillatory 
solutions are obtained in a limited range of the parameters considered 
(damping coefficient and amplitude of the high frequency force) 
for a fixed frequency and amplitude of the low frequency periodic force depending on the initial conditions.
 These solutions have distinct response amplitude to the low frequency force. 
When damping is gradually increased, only one oscillatory solution is observed. 
Vibrational resonance is observed in both the regions
 of oscillation.
	The analytical approximation yields only one oscillatory solution for all 
damping values. Analytically, the peak in the area bounded 
 by the phase portrait as a function of the amplitude of the high frequency 
	force is connected to vibrational resonance.
	Also, the values of the frequency of the low frequency forcing and 
the amplitude of the high frequency forcing at which vibrational resonance is
found to occur are obtained. In the overdamped case, vibrational resonance 
is not observed for the softening Duffing oscillator thus showing a marked 
contrast to the overdamped bistable oscillator.  

\end{abstract}

\pacs{46.40.Ff; 05.45.-a; 02.70.-c;}

\maketitle
\section{Introduction}

In conventional resonance the response of a system becomes optimum when the frequency of the input signal becomes comparable with the natural frequency of oscillation of the system. In the past few decades, different types of resonance such as chaotic resonance \cite{CR}, ghost-vibrational resonance \cite{GVR}, coherence resonance \cite{CR1}, stochastic resonance (SR), vibrational resonance (VR), etc, has been studied by changing other parameters instead of the frequency of the input periodic signal. In this work, we study the phenomenon of VR which occurs when the system is driven by two periodic forces of different frequencies. VR closely resembles the phenomenon of SR which has been found to occur in many systems \cite{SR1,SR2,SR3,SR4,SR5,SR6,SR7,SR8}. In the case of SR, the input periodic signal is added to a noisy environment. The response of the system shows peaking behavior when the noise strength is at an optimum value. However, in the case of VR, the response of the system to a low-frequency periodic force shows peaking behavior when the amplitude of a high frequency force is varied. In VR, the high frequency signal takes the place of the noise term of SR.

VR was first reported numerically in \cite{Landa} for both the 
overdamped and the underdamped bistable oscillator. The first experimental 
evidence of VR was obtained in an analog electronic circuit designed to 
model the overdamped bistable oscillator \cite{Baltanas}. The first analytical 
treatment of VR was carried out for the underdamped bistable oscillator 
\cite{Gitterman}. Over the course of the last two decades much work has been 
devoted in the study of VR. VR has been reported in various systems. For example, it occurs 
in electronic circuits based on Chua's diode and also in the FitzHugh-Nagumo model 
\cite{Ullner}, in systems with multiplicative noise \cite{Zaikin} and additive 
noise \cite{Pascual}, in the asymmetric Duffing oscillator \cite{Raj1}, systems 
with feedforward network \cite{Qin}, multistable systems with time delay \cite{Yang1} and without time delay \cite{Rajasekar}, 
fractional order potential systems \cite{Yang2}, quintic oscillators \cite{Raj2}, 
in asymmetrical deformable potentials \cite{Vincent}, in a vertical cavity surface
 emitting laser \cite{Chiz1}, in biharmonically driven plasma systems \cite{Layinde}, 
in an inhomogeneous medium with periodic dissipation \cite{Layinde1}, and in randomly connected neural networks \cite{Qin1}. 
Recently, VR has been observed in a bistable van der Pol-Mathieu-Duffing oscillator \cite{Somnath}, in an asymmetric Toda 
potential with periodic damping term \cite{Kolebaje} and in a dual-frequency-driven gyroscope excited parametrically in addition to  
an additive periodic force \cite{Oyeleke}. VR in conjunction with cascaded varying stable state non-linear systems have been found to 
detect faults in a rotating machine submerged in background noise \cite{Lei}.

In the present work, we investigate whether VR is observed in a softening 
Duffing oscillator (SDO), both theoretically as well as numerically, in the 
underdamped and overdamped regimes. 

The potential describing the SDO in the absence of damping is 
\begin{equation}
V(x) = \frac{a x^2}{2} + \frac{b x^4}{4}
\end{equation}
with $a > 0$ and $b < 0$.
This potential has one minimum at $x=0$ and two maxima at 
$x = \pm\sqrt\frac{-a}{b}$. It is a single-well with a double-hump. The well 
becomes deeper and narrower as $a$ is increased when $b = -a$. When $a$ is 
kept constant, the well becomes wider and deeper as $b \rightarrow 0$. When 
$b$ is kept constant, the well becomes wider and deeper as $a$ is increased. 
Fig. 1 shows the SDO for various $a$ and $b$ values. In our study, we fix 
the value of $a = 1$ and $b = -\frac{1}{6}$ (see inset of Fig. 1), throughout. 

\begin{figure}[htp]
\centering
\includegraphics[width=15cm,height=11cm]{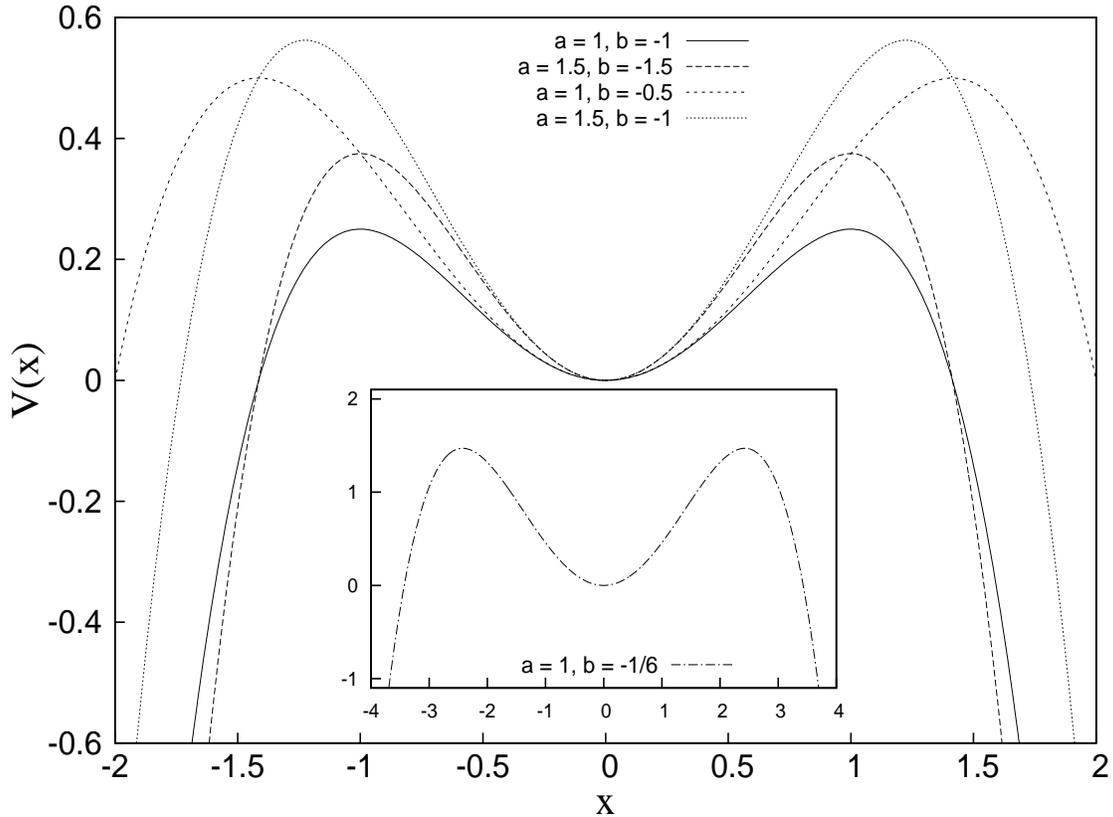}
\caption{The figure shows the SDO potential described by Eq.(1.1) for various 
$a$ and $b$ values. The inset is the SDO potential used to study in this work.}
\end{figure}

The SDO is a paradigm for many non-linear systems \cite{Kovacic}. For example, 
a harmonically excited pendulum of length $l$ whose potential is truncated upto 
third order in angular displacement $\theta$ is described by the SDO. To the best 
of our knowledge, VR in the SDO has not yet been studied extensively. In
\cite{Jeyakumari} the authors concluded analytically that for
the underdamped SDO atmost one VR is possible and multi-VR is excluded. 
We draw inspiration from their work and extend the study further.

The outline of this paper is as follows. In Sec. II, the equation of motion for 
the underdamped case is formed. Using the method of direct separation of motions 
\cite{Landa1,Baltanas}, the theoretical approximation for the response amplitude 
$Q$ of the slow motion is obtained. The response amplitude is the quantifier of VR. 
In a certain range of the amplitude $g$ of the high-frequency force and the 
damping coefficient $\gamma$, we numerically show that two locked solutions (oscillatory) 
of the underdamped SDO are obtained in addition to the unbound solution. 
The two locked solutions lead to two different values of $Q$. As a result of this, 
ensemble averaging for $Q$ is carried out for the numerical procedure. The analytical
predictions are compared with the numerical calculations. We also present a plausible analytical 
explanation for the occurrence of VR. In Sec. III, the analytical 
treatment and numerical results are presented for the overdamped SDO. The conclusions 
based on our study are given in Sec. IV.

\section{The underdamped case}

The equation of motion of the underdamped SDO driven by two periodic forces 
of frequencies $\Omega$ and $\omega$ with amplitudes $g$ and $f$ respectively is given by
\begin{equation}
\ddot{x} + \gamma \dot{x} + ax + b x^3 = f cos \omega t + g cos \Omega t
\end{equation}
where $\gamma$ is the damping coefficient and  $\Omega \gg \omega$.

\subsection{Theoretical description of vibrational resonance}

An approximate analytical solution of Eq. (2.1) is found by the method of direct 
separation of motions. According to this method, a solution is obtained in the form of 
\begin{equation}
x = X(t) + \Psi (t,\Omega t)
\end{equation}
where $X(t)$ describes the slow motion and $\Psi(t,\Omega t)$ is a $2\pi$ 
periodic function of time $\tau = \Omega t$ with mean zero wrt $\tau$,

\centerline{i.e. $<\Psi(t,\tau)> = \frac{1}{2\pi} \int_0^{2\pi} \Psi(t,\tau)dt = 
\overline{\Psi}(t,\tau) = 0$}

Putting Eq. (2.2) into Eq. (2.1) and averaging over one cycle of $\tau$, 
we have the following equation for $X$ and $\Psi$:

$\ddot{X} + \gamma \dot{X} + aX + bX^3 + b\overline{\Psi^3} + 3bX\overline{\Psi^2} = f cos \omega t$

$\ddot{\Psi} + \gamma \dot{\Psi} + a\Psi + b(\Psi^3 - \overline{\Psi^3}) + 3bX^2\Psi + 3bX(\Psi^2 - \overline{\Psi^2}) = g cos \Omega t$

Since $\Psi$ is a fast motion, we assume that $\ddot{\Psi} \gg \Psi,\Psi^2,\Psi^3,\dot \Psi$. 
Retaining only the term containing $\ddot\Psi$ in the LHS of the above equation, we get:

\begin{equation}
\ddot{\Psi} = g cos \Omega t
\end{equation}

The approximate solution for $\Psi$ then is:

\begin{equation}
\Psi  \approx -\frac{g}{\Omega^2} cos \Omega t
\end{equation}

\centerline{So, $\Psi^2 = \frac{g^2}{\Omega^4} cos^2 \Omega t$, $\overline{\Psi^2} = \frac{g^2}{2\Omega^4}$ and $\overline{\Psi^3} = 0.$}

The equation of motion for $X$ thus reduces to:

\begin{equation}
\ddot{X} + \gamma\dot{X} + aX + bX^3 + \frac{3bg^2X}{2\Omega^4} = fcos\omega t
\end{equation}

Putting $a=1$ and $b=-\frac{1}{6}$, the above equation becomes:

\begin{equation}
\ddot{X} + \gamma\dot{X} + \left(1 - \frac{g^2}{4\Omega^4}\right) X - \frac{1}{6}X^3 = fcos\omega t
\end{equation}

The effective potential $V_{eff}(X)$ corresponding to the slow motion is thus given by:

\begin{equation}
V_{eff}(X) = \left(1 - \frac{g^2}{4\Omega^4}\right) \frac{X^2}{2} - \frac{X^4}{24}
\end{equation}

\begin{figure}[htp]
\centering
\includegraphics[width=15cm,height=11cm]{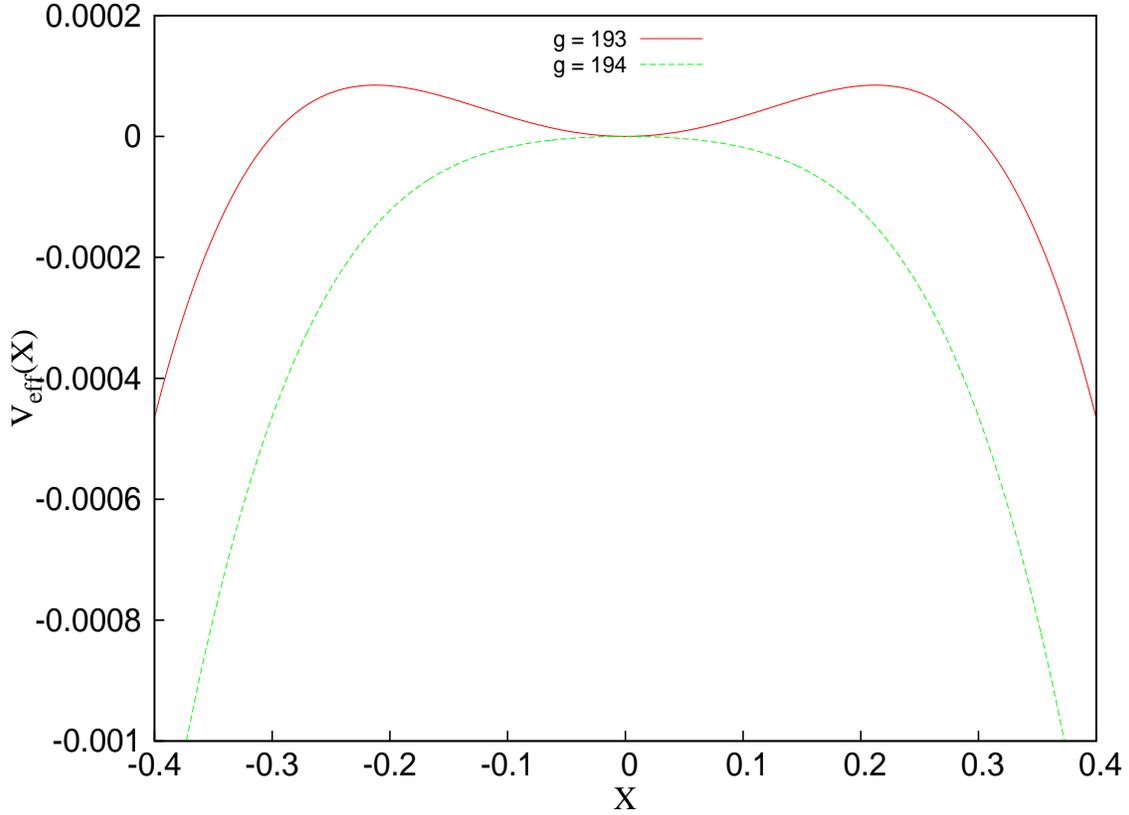}
\caption{The figure shows the effective potential $V_{eff}(X)$ described by Eq.(2.7) 
with $\Omega = 9.842$ for the $g$ values as denoted in the graph. Notice the change 
in the shape of the potential occurring at $g = g_c \approx 194.$}
\end{figure}

The shape of the effective potential gets modified by varying $g$ or $\Omega$. 
On varying $g$, $V_{eff}(X)$ is an inverted potential for $g \ge 2\Omega^2$. If 
$g < 2\Omega^2$, then $V_{eff}(X)$ is the single-well double-hump form. In 
this section we fix $\Omega = 9.842$. The change in the shape of the potential
 occurs at $g=g_c \approx 194$. Fig. 2 shows $V_{eff}(X)$ for values of $g=g_c$ and 
close to $g_c$.

From Eq. (2.7), the equilibrium points corresponding to the slow oscillations when $f=0$
can be obtained. The stable equilibrium point is $X^{*}_0 = 0$ and
the two unstable equilibrium points are $X^{*}_{(1,2)} = \pm \sqrt{6\left(1 - \frac{g^2}{4\Omega^4}\right)}$.
For this system, motion that is stable is restricted only about the 
equilibrium point $X^{*}_0 = 0$.

Let the deviation about this equilibrium point be denoted by $Y = X - X^{*}_0$.
Assuming the deviation to be small (since $f \ll 1$ and for $t \rightarrow \infty$), 
linearizing Eq. (2.6) we get:

\begin{equation}
\ddot{Y} + \gamma \dot{Y} + \omega^{2}_r Y = f cos\omega t
\end{equation}

where,
\begin{equation}
\omega_r = \sqrt{1 - \frac{g^2}{4\Omega^4}}
\end{equation}
is the resonant frequency.

On solving Eq. (2.8) by the complex exponential method, we obtained
the amplitude of the slow oscillation to be 
\begin{equation}
Y_{slow} = \frac{f}{\sqrt{{(\omega^{2}_r - \omega^2})^2 + \gamma^2 \omega^2}}
\end{equation}

The response amplitude $Q$ is the ratio of the amplitude of the 
slow oscillation to the amplitude $f$ of the small frequency force.
It is independent of $f$.

\begin{equation}
Q = \frac{Y_{slow}}{f} = \frac{1}{\sqrt{S}} =  \frac{1}{\sqrt{{(\omega^{2}_r - \omega^2})^2 + \gamma^2 \omega^2}}
\end{equation}

Now,
\begin{equation}
S = \left(1 - \frac{g^2}{4\Omega^4} - \omega^2 \right)^2 + \gamma^2 \omega^2
\end{equation}

When $S$ is a minimum, $Q$ is a maximum which corresponds to resonance.
 Below we find the variation 
of $S$ when $g$ and $\omega$ are varied.

\begin{enumerate}[(i)]
\item{The change in $S$ when $g$ is varied is

\centerline{$\frac{\partial S}{\partial g} = S_g = \frac{g}{\Omega^4}(\omega^2 +\frac{g^2}{4\Omega^4} - 1)$}

\centerline{Also, $\frac{\partial^2 S}{\partial g^2} = S_{gg} = \frac{1}{\Omega^4}\left( \omega^2 + \frac{g^2}{4\Omega^4} -1 \right) + \frac{g^2}{2\Omega^8}$}

For minima or maxima, $S_g = 0$. This occurs for $g = 2\Omega^2 \sqrt{1-\omega^2} = g_{m}$. 

And, $S_{gg} \left.\right|_{g=g_{m}} = \frac{2(1-\omega^2)}{\Omega^4} > 0$ provided $\omega^2 <1$.}

So, 
\begin{equation}
g = g_m = g_{VR} = 2\Omega^2 \sqrt{1-\omega^2} 
\end{equation}
gives the value of $g$ where resonance occurs and is independent of $\gamma$.

\item{Similarly, the change in $S$ when $\omega$ is varied is 

\centerline{$S_{\omega} = 2\omega\left[\gamma^2 - 2\left(1-\frac{g^2}{4\Omega^4} - \omega^2\right) \right]$}

}
with 
\begin{equation}
\omega_{VR} = \sqrt{1 - \frac{g^2}{4\Omega^4} - \frac{\gamma^2}{2}} = \sqrt{\omega_r^2 - \frac{\gamma^2}{2}},
\hspace{1cm}\omega_r^2 > \frac{\gamma^2}{2}
\end{equation}

\end{enumerate}

\subsection{Numerical Results}

\begin{figure}[htp]
\centering
\includegraphics[width=15cm,height=11cm]{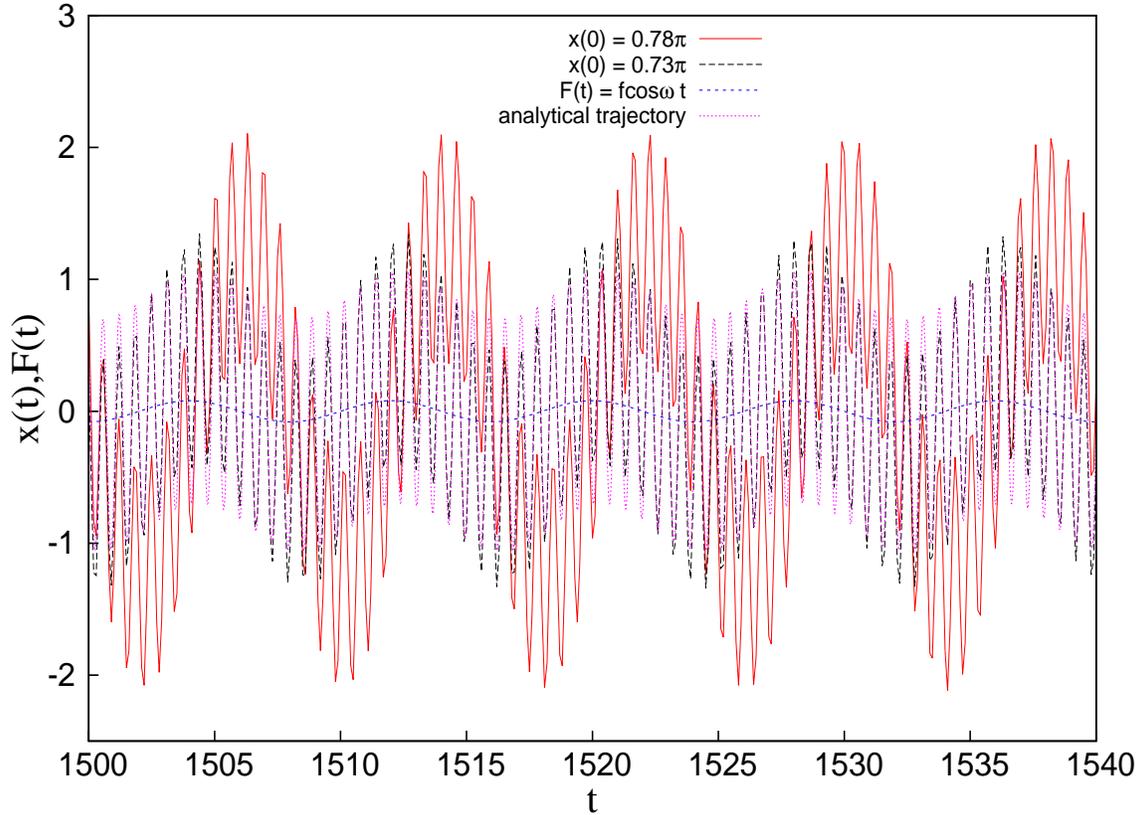}
\caption{The figure shows the trajectories $x(t)$ versus $t$. The trajectories
obtained numerically with initial positions $x(0)$ as indicated in the graph are
plotted in red and black. The trajectory obtained analytically is in pink. 
The small frequency force $F(t)$ is in blue.
Here, $f = 0.08, g = 85,\Omega = 9.842, \omega = 0.786, \gamma = 0.08$.}
\end{figure}

We solve numerically Eq. (2.1) by using the Heun's method to obtain 
the trajectories $x(t)$ for various initial conditions $x(0)=x(t=0)$. 
At $t=0$, 100 initial positions were taken in equispaced intervals in 
the range of $-0.78\pi \le x(0) < 0.78\pi$ of the SDO potential, each with initial
velocity $v(0) = v(t=0) = 0$. A time step of $\delta t =0.001$ has 
been used for the calculation. In addition to time-averaging, ensemble 
averaging has also been carried out for the numerical calculation. The reason 
for this is that although the analytical approximation gives one
locked (oscillatory) solution $x(t)$, the numerical calculation in addition to 
the unbound solution also yields two locked solutions in a certain range of the $(\gamma - g)$ plane. 
Fig. 3 shows the oscillatory trajectories $x(t)$ obtained analytically as well as numerically. 
The two numerical solutions have distinct amplitudes and phases with respect to 
the small frequency force. Due to the action of the two forcing frequencies, 
modulation of $x(t)$ is observed. The trajectories have a high-frequency component 
with frequency close to $\Omega$ and a small-frequency component (envelope) with 
a frequency close to $\omega$. The envelope of one of the numerically calculated 
trajectory has a smaller amplitude (SA) in comparison to the other trajectory 
which has a larger amplitude (LA) envelope. In Fig. 3, the SA trajectory is depicted in black
while the LA trajectory is depicted in red. From the figure, it is seen that the 
SA trajectory is nearly in-phase with respect to the small-frequency force $F(t)$ 
(in blue) while the LA trajectory is out-of-phase with respect to $F(t)$. 
The trajectory obtained analytically (in pink) closely matches the SA trajectory.  

VR is usually quantified by the response amplitude $Q$ of the system at the 
low-frequency $\omega$. It is defined as:

\begin{equation}
Q = \frac{\sqrt{Q^2_c + Q^2_s}}{f}
\end{equation}

where
\begin{equation}
Q_c = \frac{2}{nT}\int_0^{nT}x(t) cos\omega t dt
\end{equation}

and 
\begin{equation}
Q_s = \frac{2}{nT}\int_0^{nT}x(t) sin\omega t dt
\end{equation}

where, $n = 1000$ is the total number of periods of the small-frequency force $F(t)$,

and $T = \frac{2\pi}{\omega}$ is the time-period of $F(t)$.

\begin{figure}[htp]
\centering
\includegraphics[width=15cm,height=11cm]{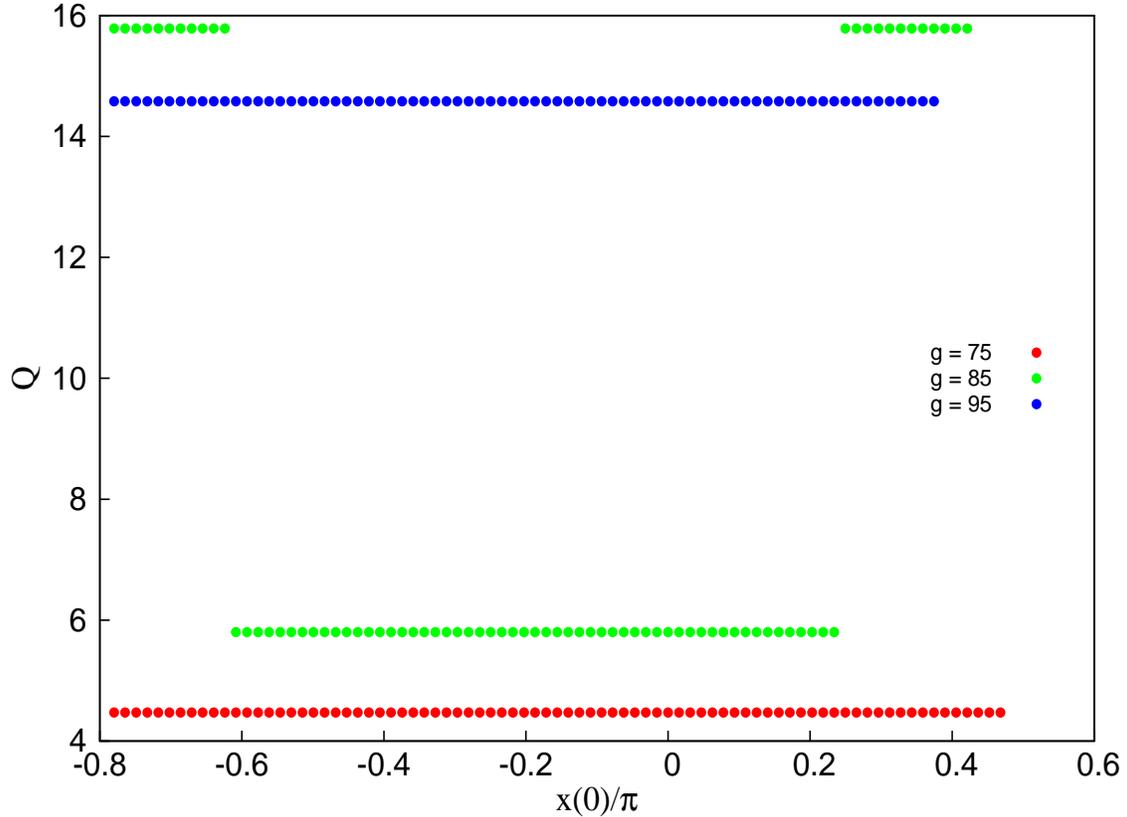}
\caption{The figure shows the values of $Q$ for different initial
 positions. Here, $f = 0.08$, $\gamma = 0.08$, $\omega = 0.786$, 
$\Omega = 9.842$ and with $g$ values as indicated in the plot.}
\end{figure}

The difference in amplitudes of the SA and LA trajectories 
result in the difference in their $Q$ values. Fig. 4 shows 
the Q values for the different initial positions $x(0)/\pi$ 
for three values of $g(= 75, 85, 95)$, with $f = 0.08$, 
$\gamma = 0.08$ and $\omega = 0.786$. It can be seen that 
for $g = 75$ and $g= 95$ there is only one value of Q for 
the different initial positions. For $g=75$ the value of Q 
is smaller in comparison to that of $g=95$. This is so because 
for the parameters chosen, for $g=75$ the locked solutions are 
in the SA state whereas for $g=95$ the locked solutions are in 
the LA state. For some initial positions, for example $0.58/\pi$, 
the solutions are unbound thus resulting in an undefined Q value 
and hence are not shown in the figure. 
For $g=85$, two bands of Q values are obtained which depend 
upon the initial positions taken. The upper band 
with $Q \approx 16$ represents the LA state whereas the lower 
band with $Q \approx 6$ represents the SA state.

\begin{figure}[htp]
\centering
\includegraphics[width=15cm,height=11cm]{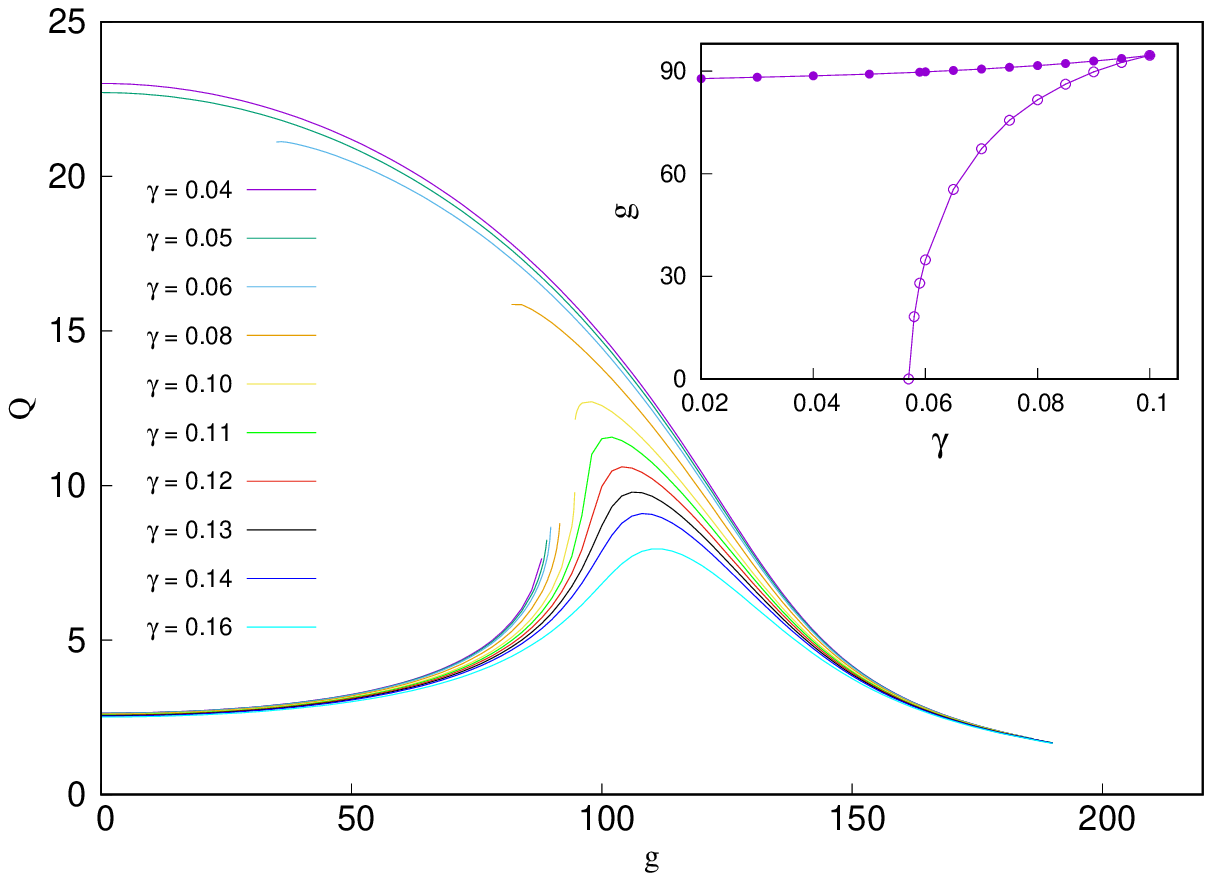}
\caption{The figure shows the variation of $Q$ as a function of $g$ for different $\gamma$ values as indicated in the plot. Here, $f = 0.08$, $\omega = 0.786$ and $\Omega = 9.842$. The inset shows the regions of existence of the SA and LA states in the $(\gamma - g)$ plane.}
\end{figure}

Fig. 5 shows a more elaborate picture to demonstrate the 
existence of the two oscillatory states. The variation of $Q$ as $g$ is increased is shown for different $\gamma$ values with $f = 0.08$,
$\gamma = 0.08$ and $\omega = 0.786$. In this figure, we choose 100 initial positions in the range $-0.78\pi\le x(0) < 0.78\pi$ and depending upon the initial position taken, the solution can either be in the LA or the SA state with a particular $Q$ value as $g$ is varied. Notice that for $\gamma = 0.04$ to $\gamma = 0.10$, two sets of $Q$ values are obtained. These two sets are not contiguous with one another. For example, for $\gamma = 0.06$, one set of $Q$ value  with $Q \approx 2.63$ begins at $g = 0$ and increases upto $Q \approx 8.67$ as $g$ is increased to $g \approx 90$. The other set begins at $g \approx 35$ with $Q \approx 21.2$ and decreases to $Q \approx 1.8$ as $g$ is increased to $g \approx 193$. The former set corresponds to the SA state while the latter set corresponds to the LA state. However, for $\gamma > 0.10$, only one continuous set of $Q$ value is obtained as $g$ is increased from $0$.

Beyond $g \approx 193$, all the trajectories obtained numerically are unbound. This value of $g$ closely match with the analytical approximation value $g_{c}$ where the effective potential $V_{eff}(X)$ corresponding to the slow motion changes to an inverted potential. 

In the inset of Fig. 5 is presented the regions of existence of the two states in the $(\gamma - g)$ plane with $f = 0.08$, $\gamma = 0.08$ and $\omega = 0.786$. The region bounded between the filled and open circles and the coordinate axes represent the parameter range where both SA and LA states are found. The region beyond the filled circles but upto $g \approx g_{c}$ and upto $\gamma \approx 0.10$ consists of only LA states and the unbound trajectories. The region to the right of the open circles but upto $\gamma \approx 0.10$ consists only of the SA states. Beyond $\gamma \approx 0.10$ but upto $g \approx g_{c}$, is the region where the two states are not distinguishable. This diagram is important in determining which set of parameter range is required  to carry out ensemble averaging. For example, if we choose $\gamma = 0.04$ then ensemble averaging is needed because for this $\gamma$ value both SA and LA states coexists from $g = 0$ upto $g \approx 90$. However, if we choose  $\gamma > 0.10$, then ensemble averaging is not necessary because only one set of stable trajectories exists with a single finite $Q$ value.

\begin{figure}[htp]
\centering
\includegraphics[width=15cm,height=11cm]{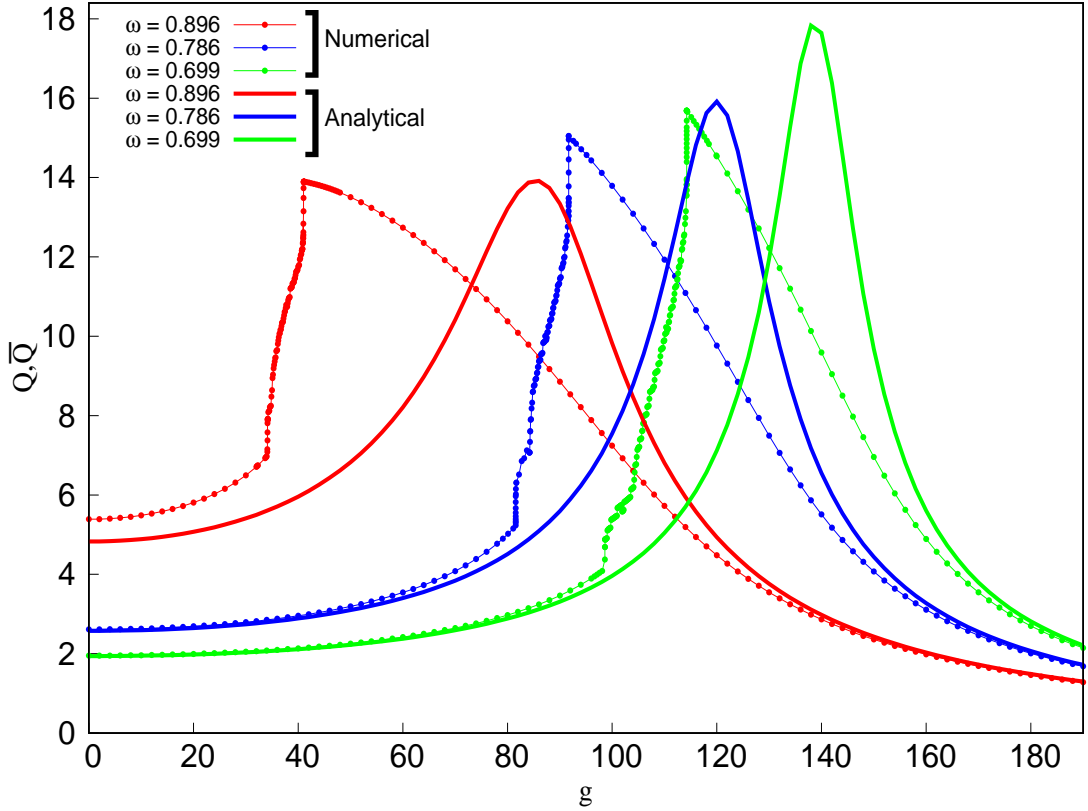}
\caption{The figure shows the variation of $Q$ (obtained analytically) and $\overline{Q}$ (obtained numerically) as a function of $g$. The analytical results are shown by thick lines while the numerical results by thin-dotted lines. Here, $\gamma = 0.08$, $f = 0.08$, $\Omega = 9.842$ for three different $\omega$ values as indicated in the plot.}
\end{figure}

Figures 6 and 7 are the main results of this section. In fig. 6, we compare the analytical result of the response amplitude $Q$ as a function of $g$ with the ensemble averaged response amplitude $\overline{Q}$ obtained numerically for three values of $\omega = (0.896, 0.786, 0.699)$. We have taken $\gamma = 0.08$ and $f = 0.08$. The analytical plots have been obtained from Eq. (2.11). As explained earlier, $\gamma = 0.08$ lies in the coexistence region of the SA and LA states, so for the numerical calculation we obtain $Q$ for each 100 initial positions in the range mentioned before and then average over the ones which give a finite $Q$ to get $\overline{Q}$. In the figure, the analytical results have been plotted with thick lines while the numerical results have been plotted by thin-dotted lines. Although quantitatively the analytical plots do not match the numerical ones but their qualitative features are similar. With an increase in $g$, $Q$ rises, peaks and then dips. This peaking behavior of $Q$ as a function of $g$ is VR. With a decrease in $\omega$, the VR peaks occur at larger $g$ values. For small $\omega$, the analytical $Q$ value closely match the numerical one at small and large $g$ values. It can be seen that the numerical plots appear wiggly as the $Q$ value begins to rise sharply. For example, for $\omega = 0.786$ the wiggles are not present when $g < 80$ and when $g > 90$ but appear only when $g$ lies in between the two. Referring to the inset of Fig. 5, it can be seen that when $g <80$ only the SA state is present and when $g > 90$ only the LA state is present, in addition to the unbound solutions. However, in $80 < g < 90$, both SA and LA states are present and their relative population changes as $g \rightarrow 90$ thus resulting in the wiggles. For $\omega = 0.896$ the region of coexistence approximately lies in the interval $34<g<40$ and for $\omega = 0.699$, the interval is $98<g<114$.

\begin{figure}[htp]
\centering
\includegraphics[width=15cm,height=11cm]{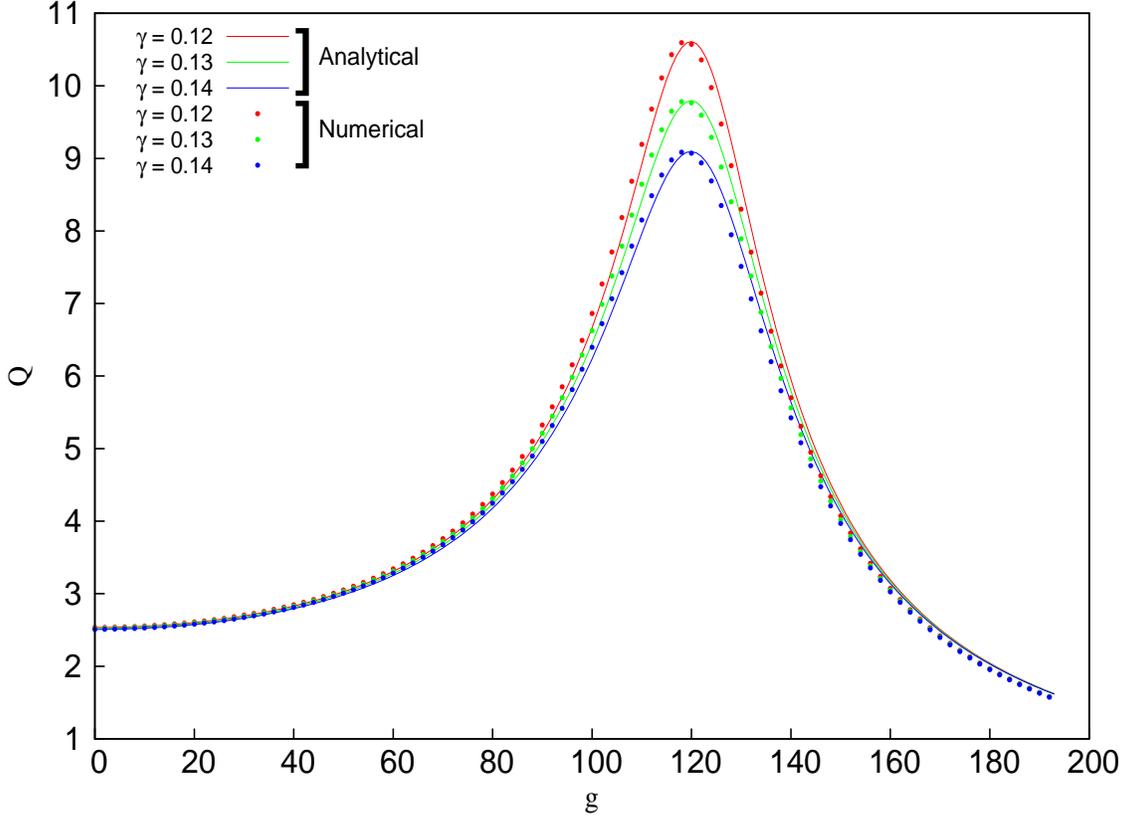}
\caption{The figure shows the variation of $Q$ as a function of $g$ for three $\gamma$ values as indicated. The analytical results are shown by lines and the numerical results by dots. Here, $\omega = 0.786$, $\Omega = 9.842$ and $f = 0.008$.}
\end{figure}

In Fig. 7, we chose $\omega = 0.786$ and $\gamma = (0.12, 0.13, 0.14)$. Clearly, as seen in Fig. 5, for these parameters taken only one stable solution is obtained. The variation of $Q$ as $g$ is increased shows VR both analytically (by lines) as well as numerically (by dots). In the figure, it is seen that the numerical result closely match the analytical result with $f = 0.008$. Notice that with an increase in $\gamma$ the peak of the $Q$ value decreases. The analytical expression for $Q$ being independent of $f$ shows a maximum at the same $g$ for the three values of $\gamma$ taken. This is because as seen in Eq. 2.13, the amplitudes $g = g_{VR}$ of the high frequency force where VR occurs, for fixed $\omega$ and $\Omega$, is independent of $\gamma$. 
 
\begin{figure}[htp]
\centering
\includegraphics[width=15cm,height=11cm]{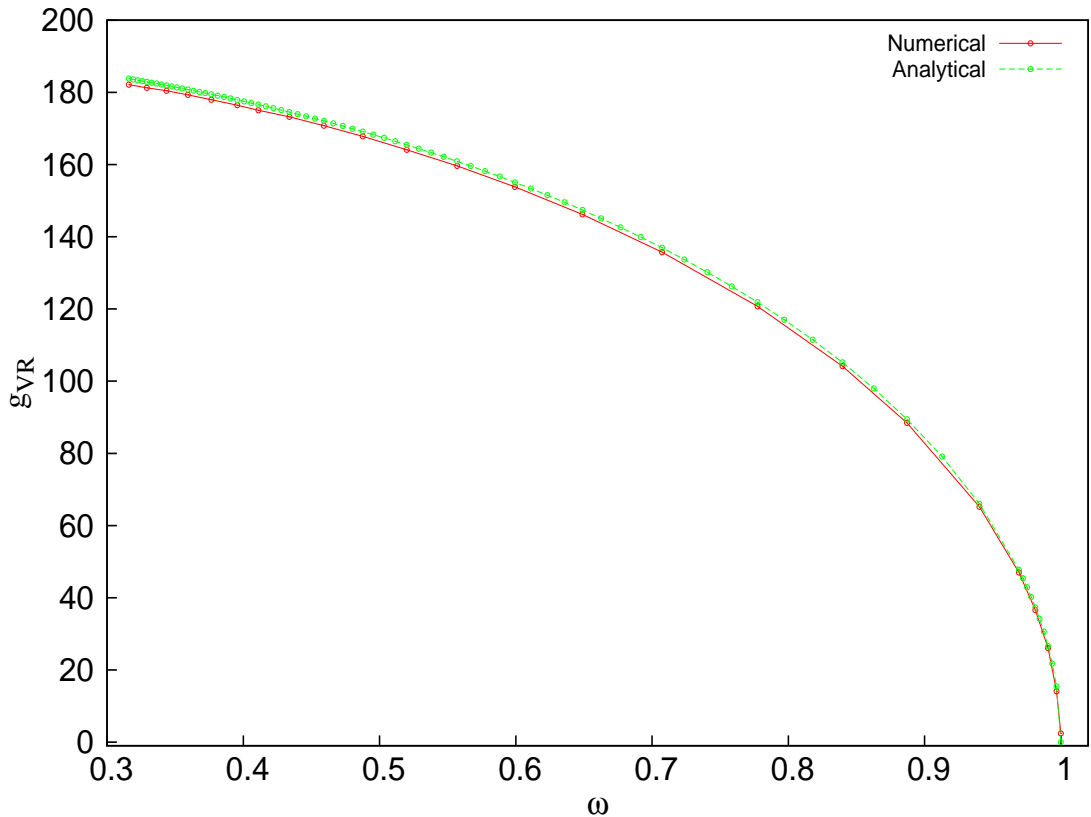}
\caption{The figure shows $g_{VR}$ as a function of $\omega$. Here, $\gamma = 0.07$ and $f=0.005$.}
\end{figure}

In Fig. 8 $g_{VR}$ is plotted as a function of $\omega$ both theoretically as well as numerically. The behavior of $g_{VR}$ shows a monotonic decrease as $\omega$ is increased. It closely matches the numerical one with parameters $\gamma = 0.07$, $f = 0.005$ and $\Omega = 9.842$. 

\begin{figure}[htp]
\centering
\includegraphics[width=15cm,height=11cm]{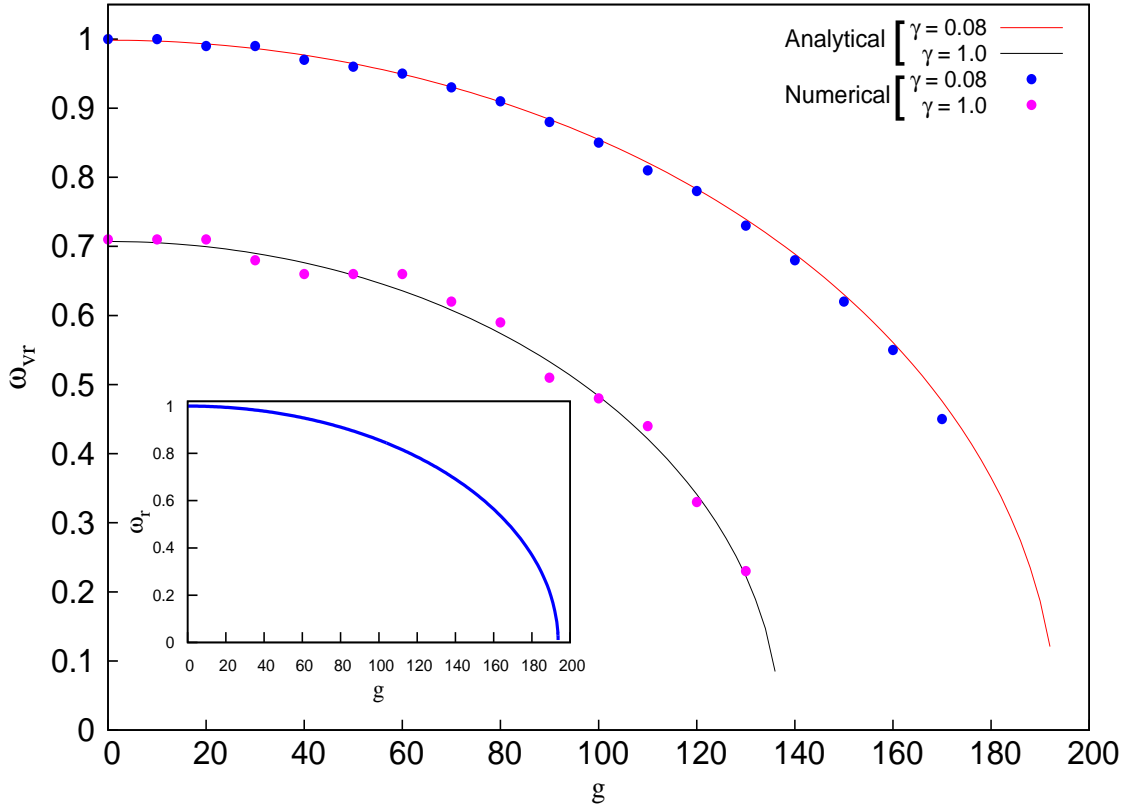}
\caption{The figure shows the frequencies $\omega_{VR}$ of the low frequency force as a function of $g$. The analytical result and numerical result has been shown for two $\gamma$ values as indicated in the graph. In the inset is plotted the resonant frequency $\omega_r$ as obtained analytically as a function of $g$.}
\end{figure}

In addition, VR is obtained for different frequencies $\omega$ of the low frequency force. Analytically, an expression for these frequencies $\omega_{VR}$ as a function of $g$ for a constant $\gamma$ has been obtained in Eq. (2.14). In Fig. 9, Eq. (2.14) is plotted for two values of $\gamma$ (in lines). It is found that for $\gamma = 0.08$ and $\gamma = 1.0$, the numerical result (in dots) is in agreement with the analytical ones for $f = 0.005$ and $f = 0.08$ respectively. The analytical expression for the resonant frequency $\omega_r$ as a function of $g$ as obtained in Eq. (2.9) is shown as an inset in Fig. 9. $\omega_r$ decreases monotonically with $g$ and becomes zero at $g \approx g_c$.

\subsection{Mechanism of VR: Analytical Explanation}

In Fig. 10, the plots marked as (a), (c), and (d) show  analytically the phase portrait of the slow motion component $X$ of the system. The figure (a) is for $\gamma = 0.08$, $f = 0.08$, $\omega = 0.786$ and for three values of $g=(100,120,140)$. 
The figure (c) is for $\gamma = 0.13$, $f = 0.008$, $\omega = 0.786$ and for the same $g$ values. Referring to Fig. 6 and Fig. 7, it is seen analytically that VR occurs for $\omega = 0.786$ at $g = g_{VR} \approx 120$. Notice that the size of the phase portraits as seen in (a) and (c) is maximum when $g = 120$. It is specifically this increase in size of the phase portrait at a certain $g$ value which leads to VR in the underdamped SDO. As $g > g_{VR}$, the size of the phase portrait gradually decreases. This trait in the phase portrait is confirmed in (b) where the area $A$ bounded by the phase portrait is obtained as a function of $g$. In (b) the peaking behavior of $A$ is seen for three values of $\omega = (0.896,0.786,0.699)$ with $\gamma = 0.08$ and $f = 0.08$. The peaks of $A$ for the three values of $\omega$ exactly coincide with the $Q$ peaks as seen in Fig. 6. This means that the area bounded by the phase portrait is a good measure of VR in this system. 
The figure (d) is for $g_{VR} \approx g = 120$, $f = 0.008$, $\omega = 0.786$ and for three values of $\gamma = (0.12,0.13,0.14)$.  The size of the phase portrait at VR is largest when the damping of the system is small. Hence in Fig.7 it is seen that the peak of $Q$ value is largest for the smallest $\gamma$ value. 

\begin{figure}[htp]
\centering
\includegraphics[width=15cm,height=11cm]{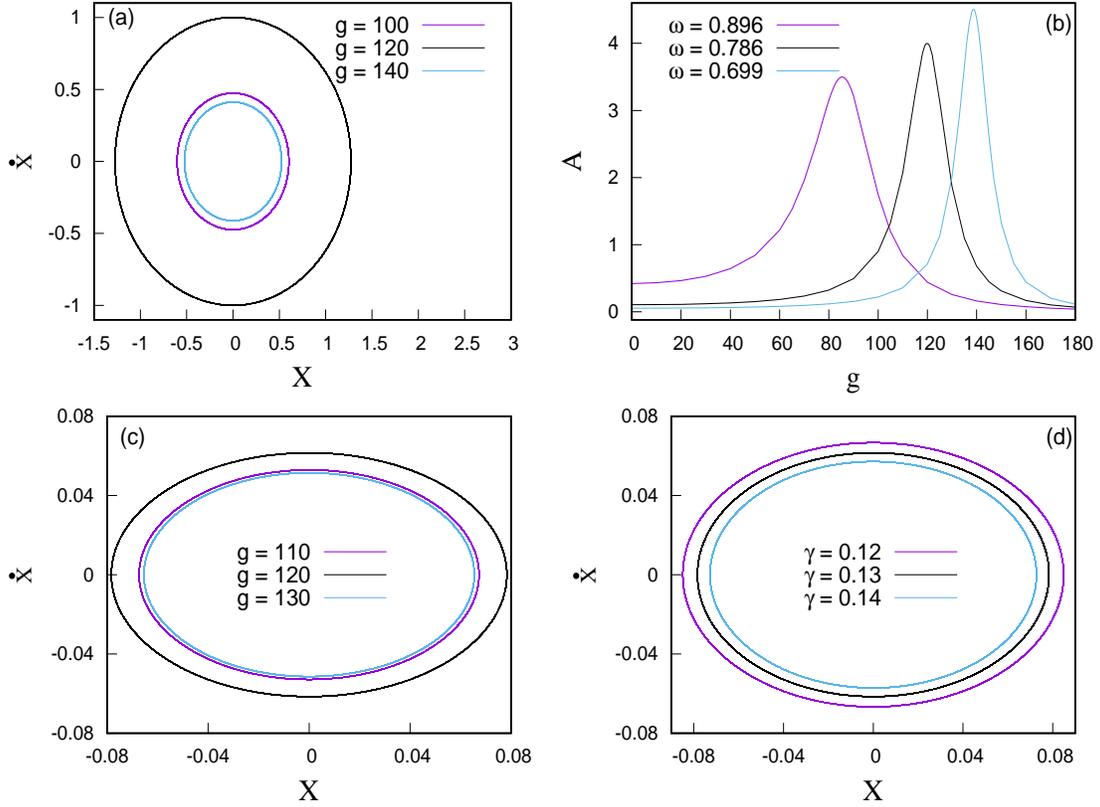}	
	\caption{The figures marked (a), (c) and (d) shows the phase portrait of the slow motion $X$ of the system as obtained analytically. The  figure (a) is for $\gamma = 0.08$, $f = 0.08$, $\omega = 0.786$ and $g$ values as indicated. The figure (c) is for $\gamma = 0.13$, $f = 0.008$, $\omega = 0.786$ and $g$ values as indicated. The figure (d) is for $g = 120$, $f = 0.008$, $\omega = 0.786$ and $\gamma$ values as indicated. The figure marked (b) shows the area $A$ bounded by the phase portrait as a function of $g$ with $\gamma = 0.08$, $f = 0.08$ for three $\omega$ values as indicated. For all the plots $\Omega = 9.842$.}
\end{figure}

\section{The overdamped case}

The equation of motion of the overdamped SDO driven by two periodic forces of frequencies $\Omega$ and $\omega$  with amplitudes $g$ and $f$ respectively is given by
\begin{equation}
\dot{x} + ax + b x^3 = f cos \omega t + g cos \Omega t
\end{equation}
with  $\Omega \gg \omega$.

\subsection{Theoretical description of vibrational resonance}

Using the same analytical method as in the underdamped case, we obtain the equations of the fast and slow motions as:

\begin{equation}
\dot{\psi} = g cos(\Omega t)
\end{equation}

and

\begin{equation}
\dot{X} + aX + bX^3 + \frac{3bg^2X}{2\Omega^2} = A cos \omega t
\end{equation}

The approximate solution for Eq. (3.2) is 
\begin{equation}
\psi \approx \frac{g}{\Omega}sin\Omega t
\end{equation}

\centerline{So, $\overline{\Psi^2} = \frac{g^2}{2\Omega^2}$ and $\overline{\Psi^3} = 0.$}

Putting the values of $a = 1$ and $b = -\frac{1}{6}$ into Eq. (3.3), we get:

\begin{equation}
\dot{X} + \left(1 - \frac{g^2}{4\Omega^2}\right) X - \frac{1}{6}X^3 = fcos\omega t
\end{equation}

The effective potential corresponding to the slow motion is
\begin{equation}
V_{eff}(X) = -\left(\frac{g^2}{4\Omega^2} - 1 \right)\frac{X^2}{2} - \frac{X^4}{4}
\end{equation}.

The stable equilibrium point is $X^{*}_0 = 0$ and
the two unstable equilibrium points are $X^{*}_{(1,2)} = \pm \sqrt{6\left(\frac{g^2}{4\Omega^2} - 1\right)}$. For $g \le 2\Omega$ the potential is the single-well double-hump form else it is inverted. In this section, we fix $\Omega = 5$. 

The slow motion is stable about the equilibrium point $X^{*}_0 = 0$. Denoting the deviation about $X^{*}_0$ as $Y = X - X^{*}_0$, and on linearizing Eq. (3.5), we get:

\begin{equation}
\dot{Y} + \omega_{r}^2 Y = f cos\omega t
\end{equation}

where the resonant frequency is:

\begin{equation}
\omega_r = \sqrt{1 - \frac{g^2}{4\Omega^2}}
\end{equation}

On solving Eq. (3.7), we obtain the response amplitude 

\begin{equation}
Q = \frac{1}{\sqrt{\omega_r^4 + \omega^2}}
\end{equation}

provided $g < 2\Omega$.

\begin{figure}[htp]
\centering
\includegraphics[width=15cm,height=11cm]{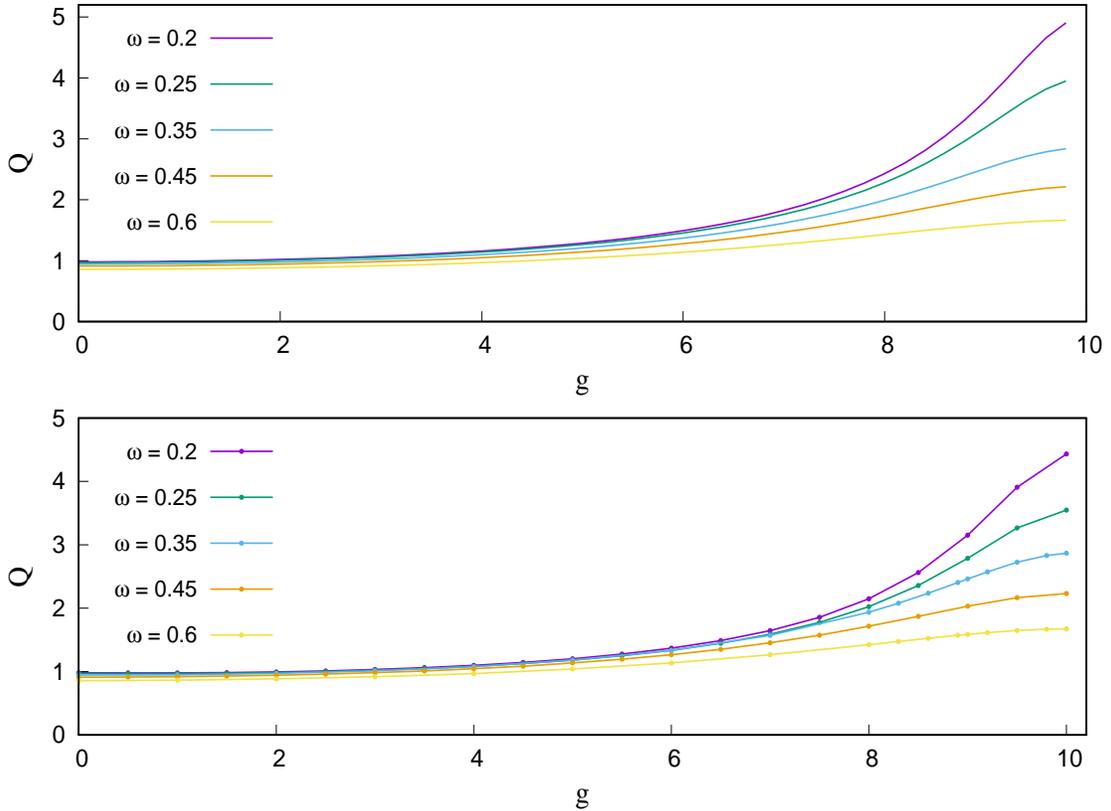}
\caption{The figure shows the response amplitude $Q$ a function of $g$. The top figure is obtained from the analytical calculation for different values of $\omega$ as indicated in the graph. The bottom figure is the one obtained numerically for the same $\omega$ values. Here, $f = 0.005$ and $\Omega = 5$.}
\end{figure}

\subsection{Numerical results}

For the overdamped SDO, the numerical result yields only one set of trajectories with finite $Q$ in addition to the unbound solution. We set $\omega < 1$ with $f = 0.005$. The top figure of Fig. 11 shows the variation of $Q$ as a function of $g$ for various values of $\omega = (0.2, 0.25, 0.35, 0.45, 0.6)$ as obtained analytically. The numerical calculation  for the variation of $Q$ as a function of $g$, for the same values of $\omega$ as above, is shown in the bottom figure of Fig. 11. It is seen that $Q$ almost remains constant as $g$ is increased before increasing monotonically with $g$ for all values of $\omega$ considered. Beyond $g >10$ all trajectories obtained numerically become unbound leading to undefined $Q$ values. This is expected because when $g > 10$ the effective potential $V_{eff}(X)$ corresponding to the slow motion changes its form to the inverted potential as shown analytically. 
For $\omega \rightarrow 1$, the analytical plots for $Q$ as $g$ is increased is in good agreement with the numerical ones. The slight differences arise when $\omega \ll 1$ for $g \rightarrow 10$. Since we observe no peaking of $Q$ as a function of $g$, we can conclude that VR is not exhibited in the case of the overdamped SDO.

\section{Conclusions}
In conclusion, this work is focussed on studying whether VR is 
exhibited or not in the SDO in both damping regimes - underdamped 
and overdamped limit. In this study, we adopted numerical procedure 
as well as a theoretical approach based on the direct separation of 
motions. This system has been driven by a periodic force of small 
frequency $\omega$ and amplitude $f$ and a periodic force of large 
frequency $\Omega$ with amplitude $g$. For both damping regimes, 
only one oscillatory solution is obtained theoretically. 
This is because the equation of motion for the slow component has
been linearized. As a consequence of this, the resulting equation 
of motion describes the forced harmonic oscillator potential in the 
presence of damping hence yielding only one steady state solution.

In the underdamped case, we numerically find the existence of two oscillatory states in a limited 
range of the $(\gamma - g)$ plane, where $\gamma$ is the damping 
coefficient. This leads to two values of the response amplitude $Q$. 
We perform ensemble averaging as well as time averaging in the set 
of parameters where the two states exist. As $\gamma$ is increased, 
only one oscillatory solution is observed for various initial conditions.
As a result, only one finite $Q$ value is obtained thereby ensemble 
averaging is not required for such a scenario.
 We compare the numerically obtained response amplitude as a 
function of the amplitude of the large frequency force with the one 
obtained theoretically. We find that VR is observed in the underdamped 
limit for both the cases discussed above. Analytically, 
the size of the phase portrait shows peaking behavior at a certain amplitude
of the high frequency force. The peak seen in the area bounded by the phase portrait 
 when the amplitude of the high frequency force is varied is a plausible 
 explanation for observing VR. In \cite{Layinde1}, the authors attributed the occurrence 
 of VR in an inhomogeneous medium with periodic dissipation to the monotonic growth 
 of the attractors.
We also obtained the values of $\omega$ and $g$ where VR is found to occur. 
The values obtained numerically are in good agreement with the theoretical predictions. 

In the overdamped limit, only one oscillatory solution is obtained numerically. We do not 
observe VR in the overdamped SDO system because as $g$ is increased beyond a certain value, 
all trajectories obtained numerically become unbound leading to undefined 
$Q$ value. Theoretically this can be explained because as $g$ is increased,
the resulting effective potential of the slow variable becomes inverted. 
It may be pointed out that VR is found to occur in a bistable oscillator, as 
cited in the introduction section, for both limits of damping. The result we 
obtained for the overdamped SDO shows another difference between the two 
non-linear oscillators.

\end{document}